\documentclass[prb,longbibliography,amssymb,twocolumn,superscriptaddress]{revtex4-1}
\usepackage{amsmath}
\usepackage{amssymb}
\usepackage{amsthm}
\usepackage{amsfonts}
\usepackage{gensymb} 
\usepackage{listings}
\usepackage{enumerate}
\usepackage{latexsym}
\usepackage{psfrag}
\usepackage{bm}
\usepackage[all]{xy}
\usepackage{graphicx}
\usepackage{subfigure}
\usepackage[pdftex,colorlinks]{hyperref}
\usepackage{color}
\usepackage{mathtools}
\usepackage{times}
\usepackage{array}
\usepackage{placeins}

\begin{document}

\title{Effective Hamiltonian for nickelate oxides
  Nd$_{1-x}$Sr$_x$NiO$_2$ 
}
\author{Hu Zhang}
\author{Lipeng Jin}

\author{Shanmin Wang}

\affiliation{Shenzhen Institute for Quantum Science and Engineering, and
  Department of Physics, Southern University of Science and Technology, Shenzhen
  518055, China}

\author{Bin Xi}
\email{xibin@yzu.edu.cn}
\affiliation{College of Physics Science and Technology, Yangzhou University, Yangzhou 225002, China}

\author{Xingqiang Shi}
\email{shixq@sustech.edu.cn}

\author{Fei Ye}  
\author{Jia-Wei Mei}
\email{meijw@sustech.edu.cn}
\affiliation{Shenzhen Institute for Quantum Science and Engineering, and Department of Physics, Southern University of Science and Technology, Shenzhen 518055, China}

\date{\today}

\begin{abstract}
We derive the effective single-band Hamiltonian in the flat NiO$_2$ planes for nickelate compounds Nd$_{1-x}$Sr$_x$NiO$_2$. We first implement the first-principles calculation to study electronic structures of nickelates using the Heyd-Scuseria-Ernzerhof hybrid density functional and derive a three-band Hubbard model for Ni-O $pd\sigma$ bands of Ni$^+$ $3d_{x^2-y^2}$ and O$^{2-}$ $2p_{x/y}$ orbitals in the NiO$_2$ planes. To obtain the effective one-band $t$-$t'$-$J$ model Hamiltonian, we perform the exact diagonalization of the three-band Hubbard model for the Ni$_5$O$_{16}$ cluster and map the low-energy spectra onto the effective one-band models. We find that the undoped NiO$_2$ plane is a Hubbard Mott insulator, and the doped holes primarily locate on Ni sites. The physics of the NiO$_2$ plane is a doped Mott insulator, described by the one-band $t$-$t'$-$J$ model with $t=265$~meV, $t'=-21$~meV and $J=28.6$~meV. We also discuss the electronic structure for the ``self-doping'' effect and heavy fermion behavior of electron pockets of Nd$^{3+}$ $5d$ character in Nd$_{1-x}$Sr$_x$NiO$_2$.
\end{abstract}

\maketitle

\section{Introduction}\label{sec:intro}
The layered high-temperature superconductors in copper oxides and iron pnictides
have motivated the search for new superconductivity compounds with layered
structures~\cite{Bednorz1986, Lee2006, Kamihara2008, Hosono2015}. Due to the
similar crystal and electronic structure, LaNiO$_2$ has been theoretical studies
as the possible analogs to the cuprates~\cite{Anisimov1999, Lee2004}. Recently,
the superconductivity with the critical temperature up to $T_c\sim 15$~K is
indeed discovered in Nd$_{0.8}$Sr$_{0.2}$NiO$_2$ thin film~\cite{Li2019}, albeit
the presence of the superconductivity remains debated~\cite{Li2019a, Zhou2019}.
Similar to copper oxides, perovskite nickelates (RNiO$_3$, where R is rare earth
or heavy metal such as Tl or Bi) display lots of strongly correlated physics
properties, such as the sharp metal-insulator transitions, particular magnetic
order, and charge order~\cite{Medarde1997, Catalan2008, Middey2016,
  Catalano2018}. The reduced form of RNiO$_3$ leads to the infinite layered
phase RNiO$_2$~\cite{Crespin1983, Hayward1999, Hayward2003, Kawai2009,
  Kaneko2009, Kawai2010, Ikeda2016}, which has a very flat NiO$_2$ plane of the
square lattice for Ni$^+$ with one hole in the $d_{x^2-y^2}$ orbital. The
superconductivity likely occurs in the NiO$_2$ plane with the charged carrier
doping.  In the cuprate superconductor compounds, the strong electronic
interactions play a significant role in the electronic
structure~\cite{Anderson1987}, and the effective one-band Hamiltonian has been
proposed to describe the low energy physics of the correlation effects for
3$d^9$ electrons~\cite{Anderson1987, Zhang1988, Hybertsen1990}. To understand
the strongly correlated electronic structures of the nickelate oxides Nd$_{1-x}$Sr$_x$NiO$_2$, we need to find out the proper effective (one-band) Hamiltonian to explore the similarity and difference from the cuprate compounds.

\begin{figure*}[t]
  \centering
  \includegraphics[width=2\columnwidth]{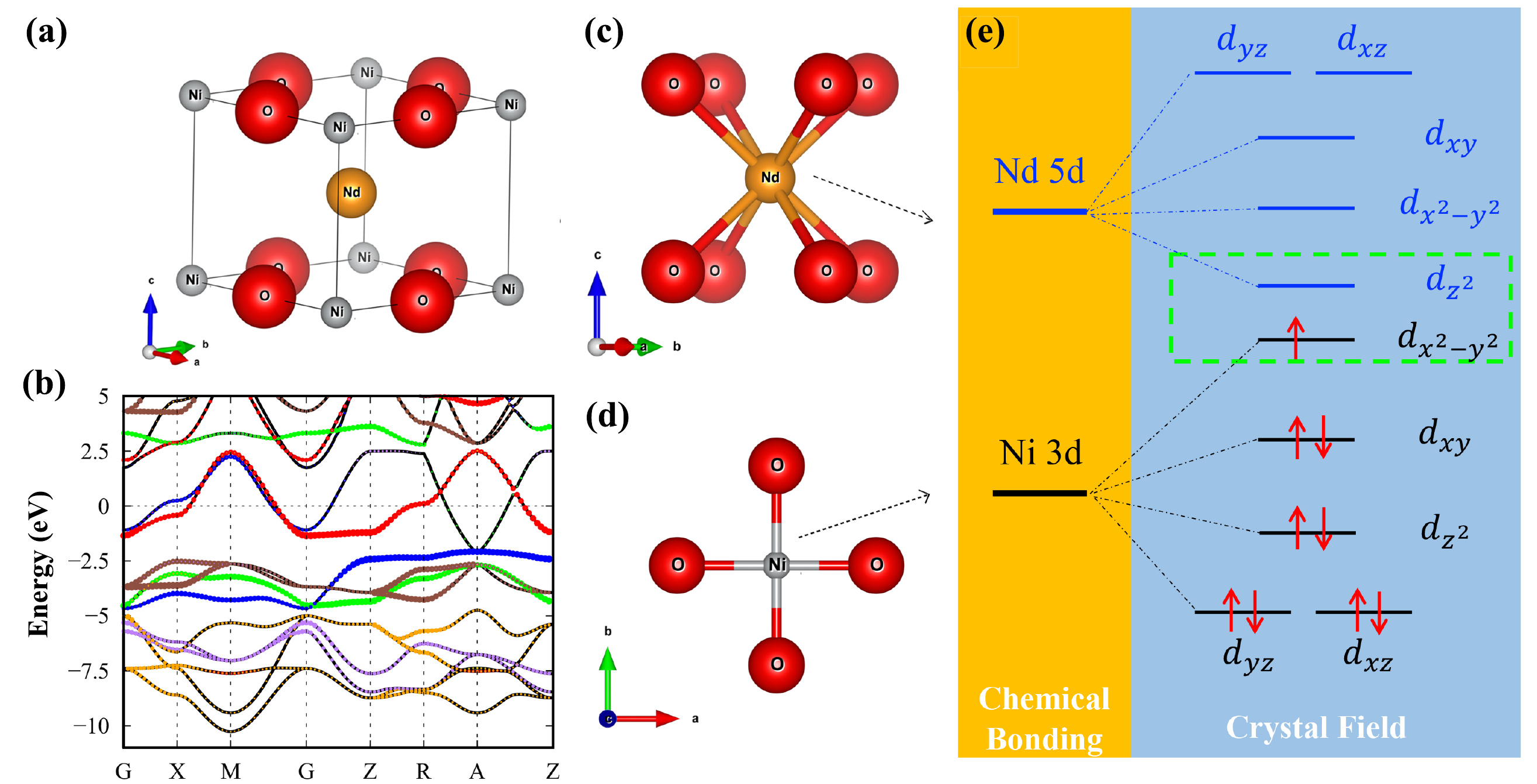} 
  \caption{(a) The crystal structure of NdNiO$_2$ with space group P4/mmm (No.
    123). (b)
    HSE06 band structures of the non-magnetic state in the parent compound
    NdNiO$_2$ along $\Gamma$(0, 0, 0)-$X$(0, 0.5, 0)-$M$(0.5, 0.5,
    0)-$\Gamma$-$Z$(0, 0, 0.5)-$R$(0, 0.5, 0.5)-$A$(0.5, 0.5, 0.5)-$Z$
    directions. The projected band structures of $d$ orbitals (red for
    $d_{x^2-y^2}$, blue for $d_{z^2}$, brown for $d_{xz/yz}$ and green for
    $d_{xy}$) in Ni and Nd, and $2p$ orbitals (orange for $p_{x/y}$ and purple
    for $p_z$) in O are also shown. The Fermi level is set at 0~eV. (c) Nd and
    its eight nearest neighbor O oxygens. (d) Ni and its four nearest
    neighbor O oxygens. (e) The crystal field splitting of Nd $5d$ and Ni $3d$ orbitals. }
  \label{fig:figure1}
\end{figure*}

NdNiO$_2$ crystallizes in the $P4/mmm$ (No.~123) space group, as depicted in
Fig.~\ref{fig:figure1}~(a). Four oxygens surround the nickel in a planar square
environment (Fig.~\ref{fig:figure1}~(d)), and the crystal field splits the $d$
orbitals as shown in Fig.~\ref{fig:figure1}~(e). Ni$^+$ has the $d^9$ electronic
state configuration, and the highest partially occupied $d$ orbital is
$3d_{x_2-y_2}$. The rare-earth ion Nd$^{3+}$ sits in the center of cuboid formed
by eight oxygen ions as shown in Fig.~\ref{fig:figure1}~(c). As Nd$^{3+}$ in
Nd$_2$CuO$_4$~\cite{Casalta1998} and Ho$^{3+}$ in
HoNiO$_3$~\cite{Fernandez-Diaz2001}, Nd$^{3+}$ has the local $4f$ moment far
below the Fermi energy level. Nd $5d$ orbitals have the split energy levels, as
shown in Fig.~\ref{fig:figure1}~(e), and near the Fermi energy, the lowest 5d
orbital in Nd$^{3+}$ is $d_{z^2}$. Therefore, in the simple reckoning for the
relevant electronic structure, there are  12 bands near Fermi energy level
corresponding to mainly Ni $d$ (5 states) and O p ($2\times3$ states) and Nd
$d_{z^2}$ (1 state). Previous density functional theory (DFT) studies within the
local density approximation (LDA) on LaNiO2~\cite{Lee2004} supports the rough
impression of the non-magnetic electronic structures.

In this paper, we first implement the first-principles simulations to derive a
three-band Hubbard model for Ni-O $pd\sigma$ bands of Ni$^+$ 3$d_{x^2-y^2}$ and
O$^{2-}$ $2p_{x/y}$ orbitals in the NiO2 planes. Based on the three-band Hubbard
model, we perform the exact diagonalization for the Ni$_5$O$_{16}$ cluster and
obtain the low-energy one-band effective Hamiltonian for the NiO$_2$ planes. We
also discuss the electronic structure for the ``self-doping'' effect and the
heavy-fermion behavior of electron pockets of Nd$^{3+}$ $5d$ character in
Nd$_{1-x}$Sr$_x$NiO$_2$. We present our main results in the two successive
stages in Sec.~\ref{sec:results}.  We discuss the physics of the one-band $t-J$
model in the NiO2 planes in Sec.~\ref{sec:discussion}. In the Appendix, we
present the results for  La$_{1-x}$Sr$_x$NiO$_2$, implying the generic
electronic structures of RNiO$_2$ series. We also include other supplementary
results for Nd$_{1-x}$Sr$_x$NiO$_2$ in the Appendix.

The main results are summarized as follows. In Sec.~\ref{subsec:DFT}, we first perform DFT
calculations of Nd$_{1-x}$Sr$_x$NiO$_2$ within the Heyd-Scuseria-Ernzerhof (HSE) hybrid
density functional. We notice that in the previous study of the nickelates
RNiO$_3$, the HSE hybrid functional method is essential to reproduce the
experimentally observed magnetic ground state~\cite{Bruno2013}.  On the
generalized gradient approximation (GGA) level, DFT simulations suggest the
G-type antiferromagnetic (AFM within the NiO$_2$ plane and AFM between NiO$_2$
planes along the $c$ direction)
ground state of the moment on Ni sites in the parent compound NdNiO2. The HSE
results for Nd$_{1-x}$Sr$_x$NiO$_2$ adopt the G-type spin configuration to mimic strong
spin correlations. In Sec.~\ref{subsec:ES}, we describe the electronic structures based on
the DFT simulations. Three-dimensional electron Fermi pockets of Nd $5d$ character
are found near Fermi energy and behave as a heavy-fermion system due to its
coupling with the local moments of Nd $4f$, similar to the electron-doped
cuprate Nd$_{1.8}$Ce$_{0.2}$CuO$_4$~\cite{Fulde1993}.  We also obtain the three-band Hubbard
model for Ni-O $pd\sigma$ bands of Ni$^+$ 3$d_{x^2-y^2}$ and O$^{2-}$ 2$p_x/p_y$ orbitals in the
NiO$_2$ planes.

In Sec.~\ref{subsec:tJ},  we derive the effective one-band Hamiltonian of the
NiO2 plane, following the standard procedure in the cuprates~\cite{Zhang1988,
  Hybertsen1990}. We derive the parameters for the three-band Hubbard model of
the Ni-O $pd\sigma$ bands for Ni$^+$ 3$d_{x^2-y^2}$ and O$^{2-}$ 2$p_{x/y}$
orbitals from the LDA results from the first-principles simulations for the
non-magnetic ground state for NdNiO$_2$.  Exact diagonalization (ED) studies of
the Ni$_5$O$_{16}$ cluster within the three-band Hubbard model are used to
select and map the low-energy spectra onto the effective one-band $t$-$t'$-$J$
model. According to ED results on finite clusters, in hole-doped nickelates, the
doped holes primarily locate on Ni sites in a good agreement with the HSE
results in Sec.\ref{subsec:DFT} and the experiment~\cite{Hepting2019}, while for
cuprate the holes mainly locate on oxygen site~\cite{Zhang1988, Hybertsen1990}.
The physics of the NiO$_2$ is a doped Mott insulator described by the effective
on-band $t$-$t'$-$J$ model with $t=265$~meV, $t'=-21$~meV, and $J=28.6$~meV. The
one-band effective Hubbard model is also given.

\section{Results}\label{sec:results}
 \subsection{DFT results of Nd$_{1-x}$Sr$_x$NiO$_2$ }\label{subsec:DFT}
 We performed first-principles calculations based on DFT~\cite{Jones2015} with the HSE06 hybrid functional~\cite{Heyd2003} as
 implemented in the Vienna Ab Initio Simulation Package (VASP)~\cite{Kresse1996,
   Gajdofmmodeheckslsesi2006, Hafner2008}. We also implement
 Perdew–Burke–Ernzerhoff (PBE) functional in generalized gradient approximation
 (GGA)~\cite{Perdew1996} and strongly constrained and appropriately normed
 semilocal density functional (SCAN) in meta-GGA~\cite{Sun2015} for comparisons.
 We use an energy cutoff of 500~eV and 12$\times$12$\times$12, 4$\times$4$\times$4, and 4$\times$4$\times$4 Monkhorst-Pack
 grids~\cite{Monkhorst1976} in the PBE, SCAN and HSE06 calculations,
 respectively. The $4f$ orbitals in Nd$^{3+}$ are expected to display the local
 magnetic moment as Nd$^{3+}$ in Nd$_2$CuO$_4$~\cite{Casalta1998} and Ho$^{3+}$ in
 HoNiO$_3$~\cite{Fernandez-Diaz2001}, and we treat them as the core-level electrons
 in the Nd pseudopotential in the HSE06 calculations. We check the results with
 4f valence electrons pseudopotential within GGA+U ($U_{4f}=10$ eV)
 scheme~\cite{Liechtenstein1995} in Appendix~\ref{sec:NdNiO}. We also check
 that the spin-orbit coupling does not significantly change the DFT results in
 the Appendix. Therefore, we only present the simulations by taking $4f$ the
 core-level electrons in Nd pseudopotential and do not include the spin-orbit
 coupling in our HSE06 hybrid functional simulations.

 We use the bulk lattice constants $a = b = 3.9208$~\AA, $c =
 3.281$~\AA~\cite{Hayward2003} for Nd$_{1-x}$Sr$_x$NiO$_2$ and didn't optimize
 the crystal structure during the DFT simulations. In this setup, the NiO$_2$
 plane remains flat in the doped case Nd$_{0.75}$Sr$_{0.25}$NiO$_2$. We didn't
 consider the lattice distortion effect upon doping in the doped nickelate
 oxides. 

Figure.~\ref{fig:figure1} (b) is the HSE06 band structure for the non-magnetic
state in NdNiO$2$. The band ordering is different from the crystal field theory
in Fig.~\ref{fig:figure1}~(e) due to the covalent effect in the hybridizations
between $d$ orbitals (Nd and Ni) and $2p$ orbitals (O). There are 12 electronic
bands near the Fermi energy $E_F$, corresponding to orbitals mainly of Ni $d$ (5
states) and O $p$ (2$\times$3 states) and Nd $d_{z^2}$ (1 state) for the
relevant electronic structure.  The O $2p$ bands extend from about -10 to -5~eV.
The Ni $3d$ bands are distributed from -5 to 2~eV, while the broad Nd $5d$
states range from -1 to 10~eV.  The Ni 3$d_{x^2-y^2}$ and Nd 5$d_{z^2}$ cross
the Fermi energy $E_F$ as expected from the crystal field splitting, as shown in
Fig.~\ref{fig:figure1}~(e). Ni 3$d_{x^2-y^2}$ is very broad along the
$\Gamma$-$X$-$M$ direction due to the strong $dp\sigma$ antibonding interaction
with oxygen $p_{x/y}$  states and encloses holes centered at the $M$ point.  Ni
3$d_{xy/xz/yz}$ bands localize near -4~eV due to the weak $dp\pi$ hybridization
with O $2p$ states. Around $A$ point, The Nd 5$d_{z^2}$ state forms an electron
pocket around the $\Gamma$ point, goes up along $\Gamma$-$Z$ direction, and
lies above the Fermi energy level with $k_z=\pi/c$. The Nd 5$d_{xy}$ lowers down
and crosses the Fermi level, forming an electron pocket around A point.

\begin{figure}[b]
  \centering
  \includegraphics[width=\columnwidth]{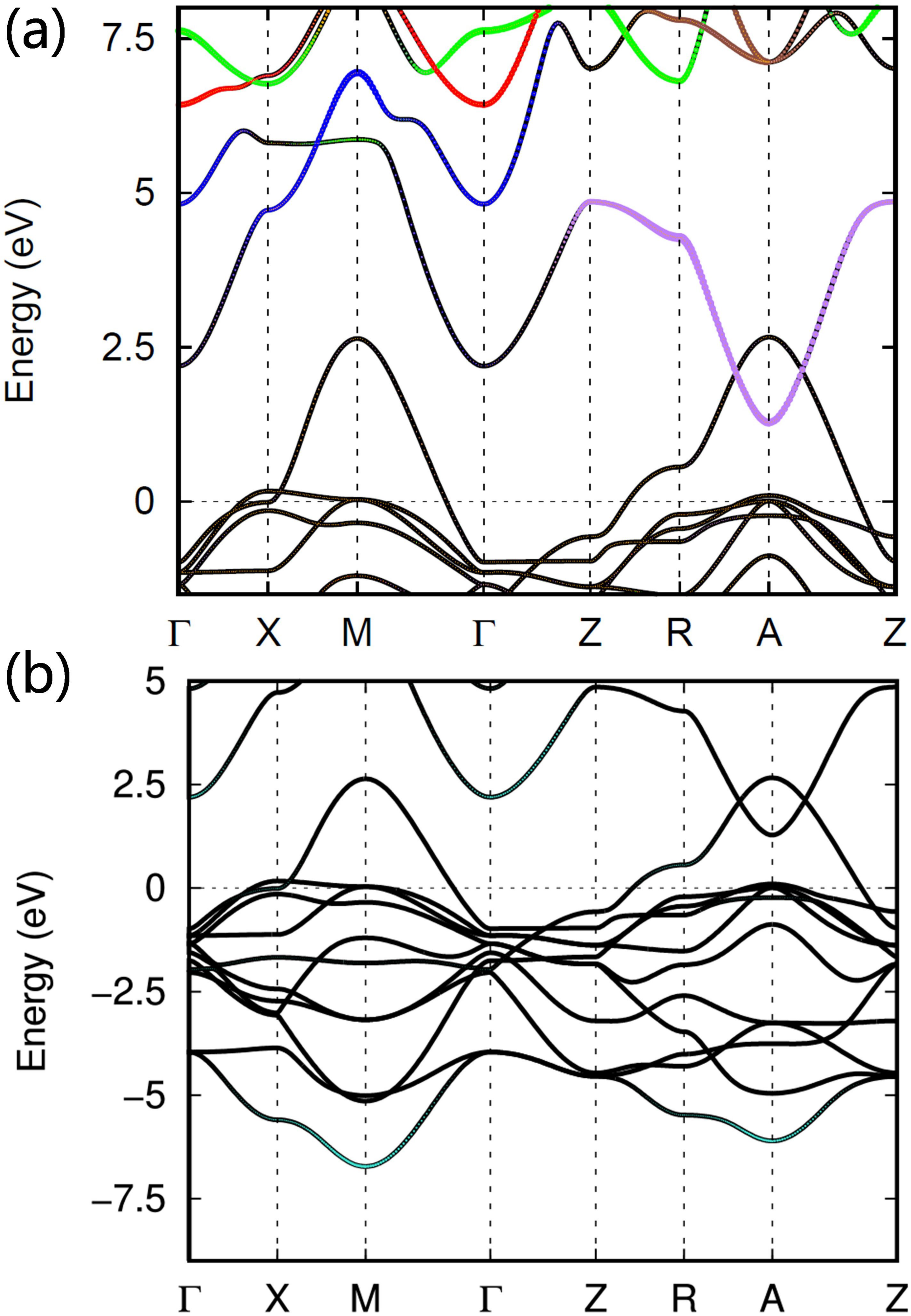}
  \caption{GGA band structures of the non-magnetic state in the simulated SrNiO2 in with the same structure as NdNiO2. The projected band structures of d orbitals in Sr, and 4p
    orbitals in Ni are also shown. The color in (a) has the same meanings as
    Fig.~\ref{fig:figure1}~(b). The color turquoise in (b) indicates the
    projected Ni 4s orbital (mainly locates at $\Gamma$ point at around 2.4~eV).}
  \label{fig:figureSrNiO2}
\end{figure}

Along $\Gamma$-$Z$ direction, four O 2$p_{x/y}$ bands and four Ni+ (3$d{xz/yz}$,
3$d_{xy}$, and 3$d_{x^2-y^2}$) bands have weak dispersions, indicating the
two-dimensional features of these bands. Ni 3$d_{z^2}$ and Nd 5d states are
dispersive along with $\Gamma$-$Z$ directions and three-dimensional extended.
The two electron pockets around $\Gamma$ and $A$ points of the Nd 5$d$ orbital
character have the Ni orbitals mixing. However, such mixing is not quickly
resolved in Fig.~\ref{fig:figure1}~(b) since the Nd 5$d$ characters dominate the
electron pockets and cover up the contributions from Ni orbitals. To further clarify the mixing
character, we also present the band structure for the simulated SrNiO$_2$ in with
the same structure as NdNiO$_2$, in order to eliminate Nd 5d orbitals in Fig.~\ref{fig:figureSrNiO2}. From the
comparing between Fig.~\ref{fig:figure1}~(b) and Fig.~\ref{fig:figureSrNiO2}, we can see that Nd 5$d_{z^2}$ band crosses $E_F$
around $\Gamma$ point with Ni 4$s$ mixing, while Nd 5$d_{xy}$ band crosses $E_F$ around A
point with Ni 4$p_z$ mixing.

The GGA band structure of NdNiO$_2$ in Appendix \ref{sec:NdNiO}  is quite similar to the LDA band
structure of LaNiO$_2$~\cite{Lee2004}, indicating a generic electronic structure in
the RNiO$_2$ family. Compared with the GGA band structure (Appendix \ref{sec:NdNiO}), for the
non-magnetic state, the mixing of the exact exchange in the HSE06 hybrid
functional separates the two bands crossing the Fermi energy $E_F$ away from
other bands, without significant change of dispersions and relative positions of
the bands far from $E_F$. The separation of two bands can also be reproduced in the
LDA+U results~\cite{Lee2004}. However, in the LDA+U scheme, Ni 3$d_{z^2}$ is raised by
U and crosses $E_F$ when U is large~\cite{Lee2004}, different from the
HSE06 hybrid functional simulation for the non-magnetic state.

 \begin{table}[b]
   \caption{\label{tab:table1} Theoretical magnetic moments (in $\mu_B$ ) on Ni in NdNiO$_2$
     calculated with HSE06, SCAN, and PBE functional. $\Delta E$ is the energy difference between the non-magnetic state and the ferromagnetic state.}
   \begin{ruledtabular}
     \begin{tabular}{ccc}
       Functional & Magnetization  & $\Delta E$ (meV)~~~   \\ 
       \hline
       HSE06& 0.94  & 744.7  \\
       SCAN& 0.76 &   142.7 \\
       GGA&  0.05 &   -0.1 \\       
     \end{tabular}
   \end{ruledtabular}
 \end{table}

 The magnetization measurement and neutron powder diffraction didn't reveal the
 long-magnetic order in LaNiO$_2$ and NdNiO$_2$; however, the paramagnetic
 susceptibilities imply the (at least short) spin correlations
 ~\cite{Hayward1999, Hayward2003}. The absence of long-range magnetic order may
 be due to poor sample qualities, or due to the ``self-doping'' effects of the
 Nd 5$d$ electron pockets. The strong correlation for electrons on Ni$^+$
 induce the magnetism in the system.

 To demonstrate the correlation and magnetism in NdNiO$_2$, we calculate the
 magnetic moment and compare the ground state with the non-magnetic one within
 different functionals (GGA, SCAN, and HSE06). To save computation time, we
 consider the ferromagnetic spin configuration in the primary unit cell. The
 results are list in TABLE~\ref{tab:table1}. Than the non-magnetic state, the magnetic states
 have increased moments on Ni with increasingly lower ground state energies from
 GGA, SCAN to HSE06. For GGA, the magnetic state has even higher energy than the
 non-magnetic state, indicating the non-magnetic ground state within the GGA
 functional, consistent with the previous study in LaNiO2~\cite{Lee2004}. The
 SCAN functional includes more correlation effects, and the magnetism is
 significantly enhanced. The magnetic ground state has further lower energy than
 the non-magnetic state in the HSE06 functional. The fact that the correlation
 achieves the magnetism is a strong indication that NdNiO$_2$ is magnetic if we
 include the correlations.

 Even though there is no long-range magnetic order, we still impose static
 magnetic configurations on Ni ions in Nd$_{1-x}$Sr$_x$NiO$_2$ to mimic the spin
 correlations in the DFT simulations for electronic structures. We find the
 G-type AFM  magnetic state has the lowest ground state energy within the GGA DFT
 simulations. Therefore, in the HSE06 simulations, we adopt the G-type AFM spin
 configurations on Ni using the $\sqrt{2}\times\sqrt{2}\times2$ supercell to study the electronic
 structures of Nd$_{1-x}$Sr$_x$NiO$_2$ ($x=0, 0.25$). Therefore, the Brillouin zone is folded
 by the G-type AFM spin configuration.  However, without any confusion, we still
 use notations, $\Gamma, X, M, Z, R, A$, for high symmetry k-points in the folded
 Brillouin zone. 

 \begin{figure}[b]
   \centering
   \includegraphics[width=\columnwidth]{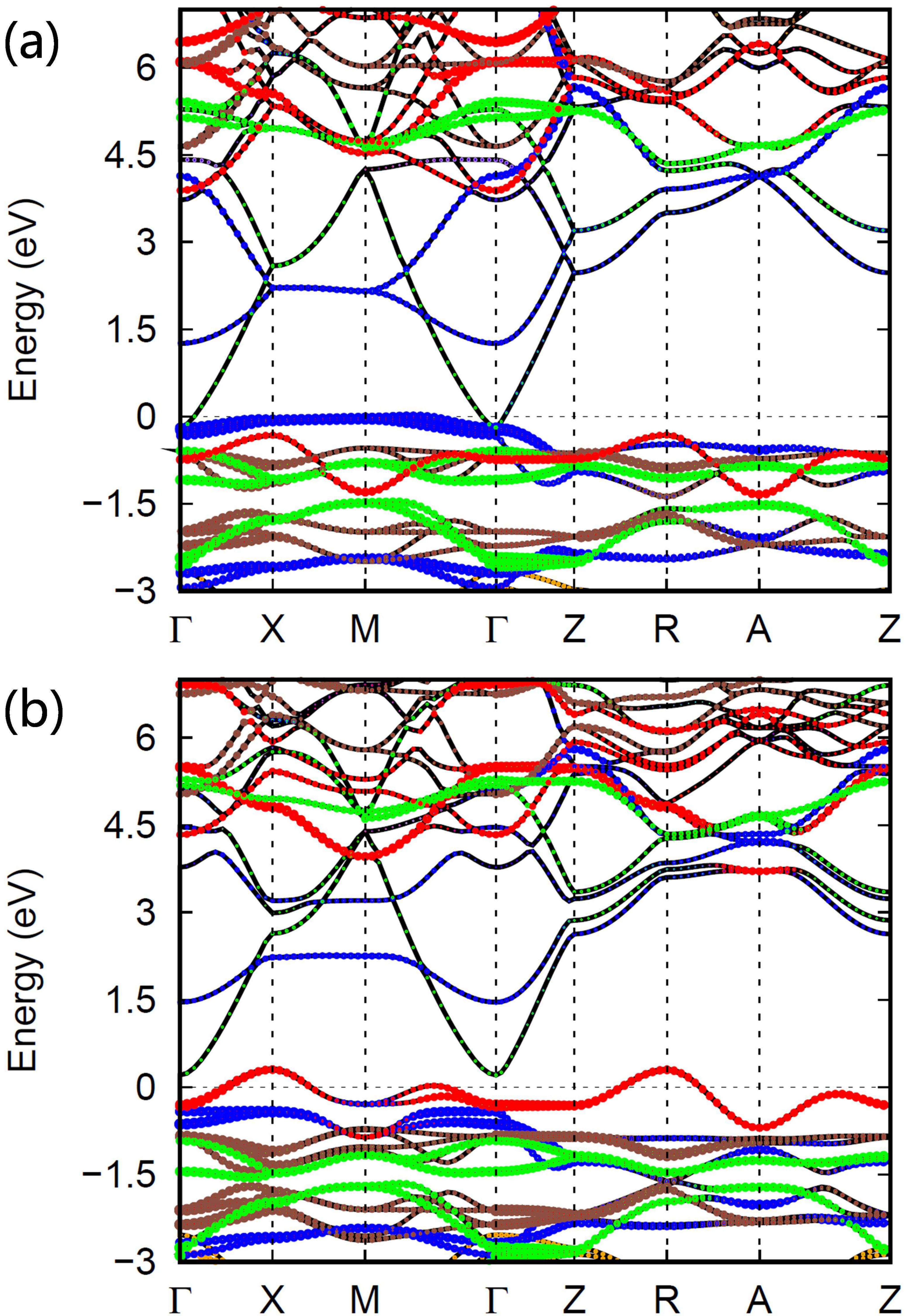}  
   \caption{HSE06 band structures of G-type AFM states for a
     $\sqrt{2}\times\sqrt{2}\times2$ supercell in (a) NdNiO$_2$ and (b)
     Nd$_{0.75}$Sr$_{0.25}$NiO$_2$. The notations of $\Gamma$, $X$, $M$, $Z$, $R$
     and $A$ are in the folded magnetic Brillouin zone. The color has the same
     meanings as Fig.~\ref{fig:figure1}~(b). 
   }
   \label{fig:band_afm}
 \end{figure}
 Figure.~\ref{fig:band_afm} displays the HSE06 band structures of
 Nd$_{1-x}$Sr$_x$NiO$_2$ ($x=0, 0.25$) with G-type AFM magnetic configuration on
 Ni.  In the HSE06 hybrid functional results, the AFM magnetic configuration not
 only folds the band structure, but also dramatically change the bands near
 the Fermi energy $E_F$, very different from the GGA result for the G-type AFM
 magnetic configuration in the Appendix~\ref{sec:NdNiO}. The significant change of band
 structures in the AFM states between GGA and HSE06 implies strong correlations
 in Nd$_{1-x}$Sr$_x$NiO$_2$.  

 In the G-type AFM state of NdNiO$_2$ (Fig.~\ref{fig:band_afm}~(a)), the Nd
 5$d_{z^2}$ band is raised above the Fermi energy $E_F$ by the correlation,
 eliminating the electron pocket of the Nd 5$d_{z^2}$ character in the
 non-magnetic state (Fig.~\ref{fig:figure1}~(b)). The Nd 5$d_{xy}$ band is also
 raised, but the electron pocket of this band still exists. We remind here again
 that the electron pocket of the Nd 5$d_{xy}$ character has the Ni $4p_z$
 orbital mixing. The Nd 5$d_{xy}$ electron pocket locates around $\Gamma$ point
 in the folded magnetic Brillouin zone. Therefore, the AFM spin correlation has
 an essential influence on the ``self-doping'' effect of Nd 5$d$ electron
 pockets. For the Ni 3$d$ band, the AFM correlation significantly renormalizes
 the bandwidth. The Ni 3$d_{x^2-y^2}$ bands split into upper and lower Hubbard
 bands separated by around 5 eV due to the AFM spin configuration, indicating
 the strong correlation in NdNiO$_2$. The lower Hubbard 3$d_{x^2-y^2}$ band
 locates lower than the Ni 3$d_{z^2}$ band, bring the doping problem into a
 complicated situation.  

When one Sr$^{2+}$ ion substitutes Nd$^{3+}$ in NdNiO$_2$, we dope an extra hole
into the NiO$_2$ planes. To simulate the doped nickelate oxide, we calculate the
band structures of Nd$_{0.75}$Sr$_{0.25}$NiO$_2$ (Nd$_3$SrNi$_4$O$_8$) in the $\sqrt{2}\times\sqrt{2}\times2$ supercell with
the G-type AFM spin configuration on Ni ions. Fig.~\ref{fig:band_afm}~(b) is the HSE06 band
structure of Nd$_{0.75}$Sr$_{0.25}$NiO$_2$. The Nd 5$d_{xy}$ band goes up above the Fermi energy
level, and the band minimum locates 0.2~eV. We can expect that Nd 5$d_{xy}$ band
contributes Hall coefficient at high temperatures at low dopings~\cite{Li2019}.
The doped hole does not go to the Ni 3$d_{z^2}$ band. Instead, it locates on the
3$d_{x^2-y^2}$ band, which goes up above 3$d_{z^2}$ in Nd$_{0.75}$Sr$_{0.25}$NiO$_2$. Therefore, the doped
hole does not polarize Ni$^{2+}$ into the $S=1$ local moment, but create an $S=0$
hole-like doped charge carrier. The Ni 3$d_{x^2-y^2}$ encloses the hole
pockets around $X$ point in the folded Brillouin zone ($M$ in the original
non-magnetic Brillouin zone). The HSE06 band structures in Fig.~\ref{fig:band_afm} suggest that
the undoped NiO$_2$ plane is Hubbard-like Mott insulator, not the
charge-transfer-like one as in the cuprate.  The doped hole goes into Ni
3$d_{x^2-y^2}$ orbital, creating an $S=0$ hole-like charge carrier, rather than
into the O$^{2-}$ 2$p_{x/y}$ orbitals. 
 
\subsection{Electronic structures of Nd$_{1-x}$Sr$_x$NiO$_2$ }\label{subsec:ES}
The presence of the electron pockets of Nd 5d bands suggests the ``self-doping''
effect, implying a small charge transfer from these pockets to the Ni-O sheets
even without chemical doping. The ``self-doping'' effect also exists in the
cuprate family, e.g., YBa$_2$Cu$_3$O$_7$ and
Bi$_2$Sr$_2$CaCu$_2$O$_8$~\cite{Pickett1989}. The three-dimensional electron
pockets of Nd 5$d_{xy}$ states have the mixing with Ni 4$p_z$ orbitals. The
``self-doping'' effect allows some charge transfer and changes the hole count in
the NiO$_2$ planes, resulting in the metallic behavior even without chemical
doping. In the Appendix~\ref{sec:NdNiO}, we present the GGA band structure with
the 4$f$ valence electron pseudopotential within the GGA+U scheme. We can see
that Nd 4$f$ states have the mixing with Nd 5$d$ states. Due to the presence of Nd
4$f$ orbitals, the electron pockets of the Nd 5$d_{xy}$ character hybridizes with the
4$f$ local moments, behaving as a heavy-fermion system 
\begin{eqnarray}
  \label{eq:5d4f}
  H_{K}=\sum_{\mathbf{k}}\epsilon_{\mathbf{k}} a_{\mathbf{k}\sigma}^\dag
  a_{\mathbf{k}\sigma}+J_K\sum_{i}\mathbf{s}_{i}\cdot \mathbf{S}_i^{4f},
\end{eqnarray}
where $a_{\mathbf{k}\sigma}$ creates fermion with momentum $\mathbf{k}$ on the electron pockets and $\mathbf{s}_i$
and $\mathbf{S}_i^{4f}$ are the spin operators for the $5d$ electrons and $4f$
local moments, respectively.


\begin{figure}[b]
  \centering    
  \includegraphics[width=0.89\columnwidth]{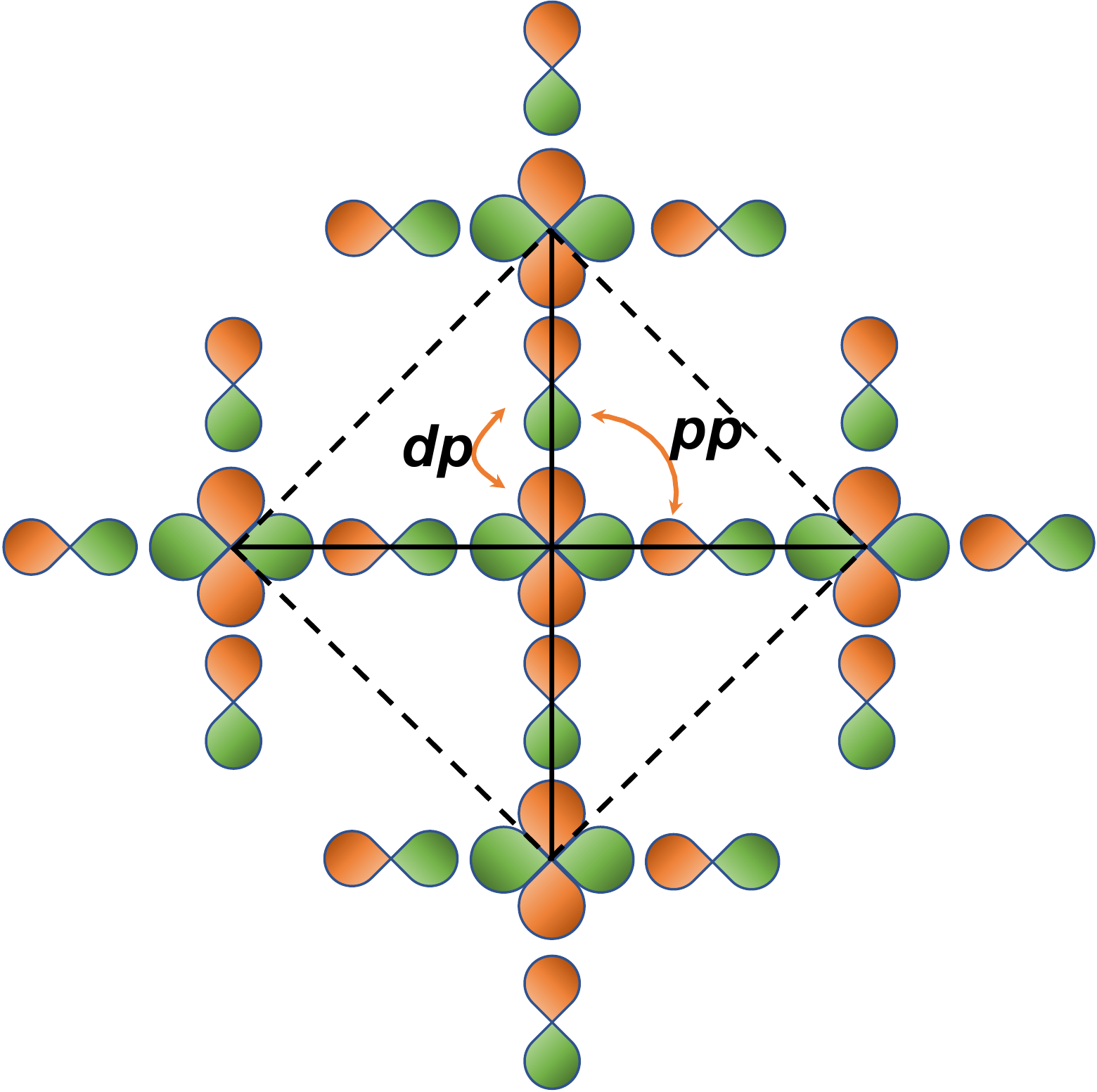}  
  \caption{Schematic representation of the orbitals (Ni $3d_{x^2-y^2}$ and O $p_{x/y}$) included in the three-band
    Hubbard model in the Ni$_5$O$_{16}$ cluster. The effective one-band model
    has the degree of freedom only on the five Ni $3d_{x^2-y^2}$ orbitals connected by the
    solid ($t$ and $J$) and dashed ($t'$ and $J'$) lines.}
  \label{fig:Ni5O16}
\end{figure}

We now turn to the physics of the NiO$_2$ plane with the hole doping, following
the process for the CuO$_2$ planes in cuprates~\cite{Zhang1988, Hybertsen1990}.
The HSE06 band structures in Fig.~\ref{fig:band_afm} suggest that the undoped NiO$_2$ plane is
Hubbard-like Mott insulator, and the doped hole goes into Ni 3$d_{x^2-y^2}$ orbital,
creating an $S=0$ hole-like charge carrier. Therefore, the physics of the NiO$_2$ plane with the
charge doping is described by the three-band Hubbard model for the $dp\sigma$
bands of Ni 3$d_{x^2-y^2}$ and O 2$p_{x/y}$ orbitals. Fig.~\ref{fig:Ni5O16} schematically presents the
orbitals of Ni 3$d_{x^2-y^2}$ and O 2$p_{x/y}$ with the green and brown colors corresponding
to the positive and negative signs of wave functions, respectively.

We assume a vacuum state $d^{10}p^6$ and introduce the operators
$d_{i\sigma}^\dag$ and $p_{l\sigma}^\dag$ creating the Ni
3$d_{x^2-y^2}$ hole and the O 2$p_{x/y}$ hole at the $i$-the Ni site and the
$l$-th O site, respectively, with the spin $\sigma=\uparrow/\downarrow$. The
holes hop between Ni  3$d_{x^2-y^2}$ and O 2$p_{x/y}$ orbitals with the amplitude $t_{dp}$, and
among O 2$p_{x/y}$ orbitals with the amplitude $t_{pp}$.  We set 
the on-site potential of Ni
3$d_{x^2-y^2}$ as $\epsilon_d=0$, and the chemical potential difference between Ni 3$d_{x^2-y^2}$
and O 2$p_{x/y}$ orbitals as $\epsilon=\epsilon_p-\epsilon_d$.  The strong
correlations involve the on-site interactions $U_p$ and $U_d$, and inter-site
interactions $U_{dp}$ and $U_{pp}$ for holes on O 2$p_{x/y}$ and Ni 3$d_{x^2-y^2}$ orbitals. The
three-band Hubbard model of the dpsigma bands reads out 
 \begin{eqnarray}
   \label{eq:3band}
   H_{dp}&=&\sum_{\langle{il}\rangle\sigma}t_{dp}(d_{i\sigma}^\dag
        p_{l\sigma}+\text{h.c.})+\sum_{\langle ll' \rangle\sigma}t_{pp}(p_{l\sigma}^\dag p_{l'\sigma}+\text{h.c.})\nonumber\\
    &+&\sum_{l\sigma}\epsilon p_{l\sigma}^\dag
        p_{l\sigma}+U_d\sum_in_{di\uparrow}n_{di\downarrow}+U_p\sum_{l} n_{l\uparrow}n_{l\downarrow}\nonumber\\
    &+&\sum_{\langle il\rangle}U_{dp}n_{di}n_{pl}+\sum_{\langle  ll' \rangle}U_{pp}n_{pl}n_{pl'}.
 \end{eqnarray}
 Here $\langle \cdots \rangle$ denotes the nearest neighbor bonds.

\subsection{Effective $t$-$t'$-$J$ model Hamiltonian of NiO$_2$ planes}\label{subsec:tJ}
\begin{figure}[b]
  \centering    
  \includegraphics[width=\columnwidth]{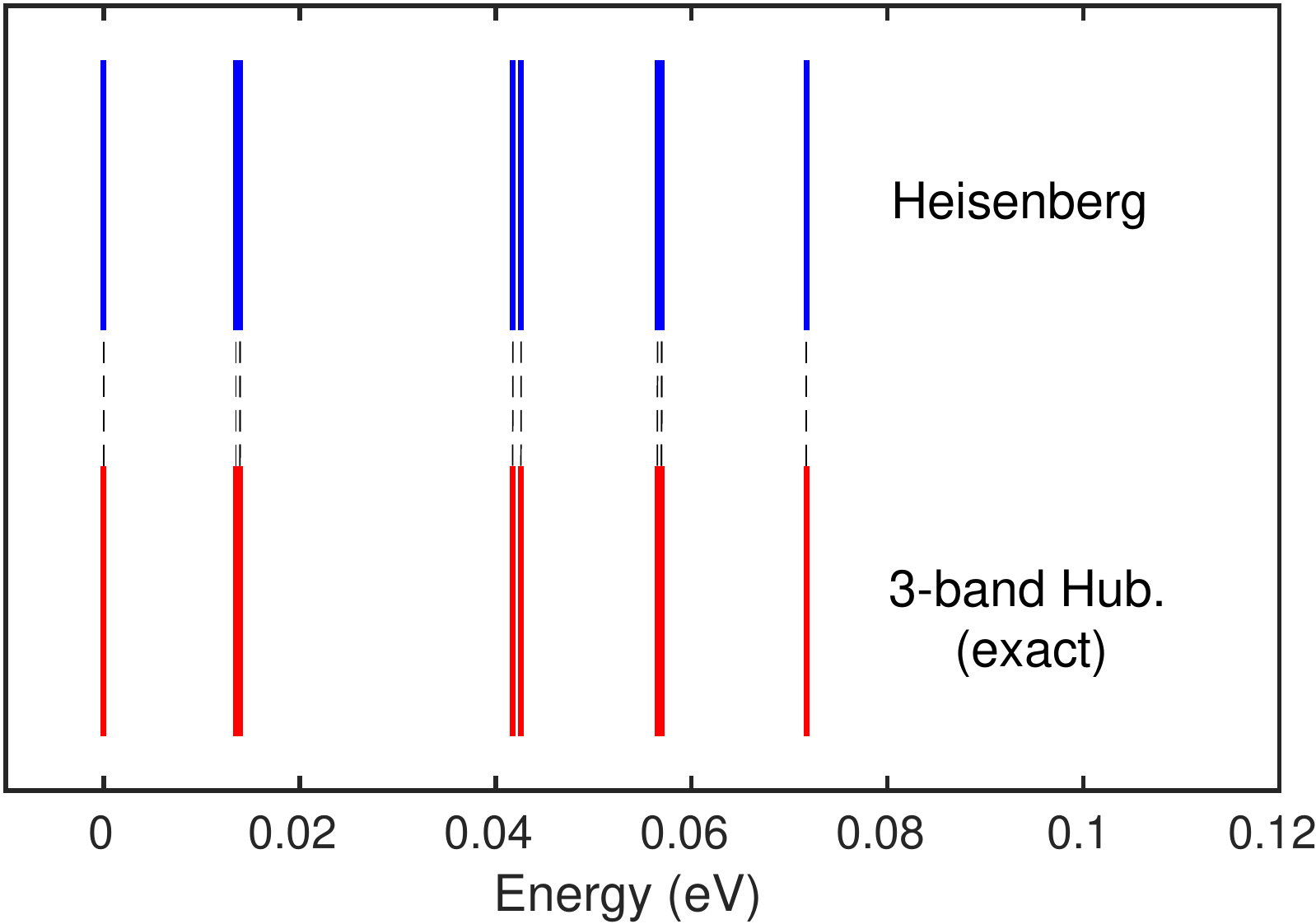}  
  \caption{The low-energy spectrum for Ni$_5$O$_{16}$ cluster calculated in the
    three-band Hubbard model in comparision to the mappings onto the Heisenberg
    Hamiltonian for the five-hole case.    }  
  \label{fig:Heisenberg}
\end{figure}
The tight-binding parameters $t_{dp}$, $t_{pp}$, and $\epsilon$ can be obtained
from the Wannier fitting~\cite{Mostofi2008, Mostofi2014} of the 12 bands in the
LDA simulations for the non-magnetic band structure of NdNiO$_2$. It is
noteworthy that the HSE06 band structure in Fig.~\ref{fig:figure1}~(b) already
contains the renormalization effect due to the strong correlations. There is
double-counting for the correlations if we obtain the tight-binding parameters
from the HSE06 band structure. The parameters are given as $\epsilon=4.2$~eV,
$t_{pd}=1.3$~eV, $t_{pp}=0.6$~eV. We choose a proper basis of the $3d_{x^2-y^2}$
and $2p_{x/y}$ for all positive $t_{pd}$ and $t_{pp}$. To match the similar
sizable band splitting of the lower and upper Hubbard bands of the Ni 3$d_{x^2-y^2}$
orbitals in the HSE06 band structure in Fig~\ref{fig:band_afm}, we set the interaction terms as
$U_d=7.5$~eV, $U_p = 5.0$~eV, $U_{pd} = 4.0$~eV, and $U_{pp} = 2.0$~eV in the ED calculation for
the derivation of the effective one-band Hamiltonian. 

We follow the process in the cuprates~\cite{Zhang1988,Hybertsen1990} to obtain
the effective one-band Hamiltonian of the NiO$_2$ planes. We perform direct ED
studies of the three-band Hubbard model $H_{dp}$ in Eq.~\ref{eq:3band} and find the effective
one-band Hamiltonian by the low-energy spectrum mapping. We carry out the ED
calculation for the Ni$_5$O$_{16}$ cluster as shown in Fig.~\ref{fig:Ni5O16} with five and six holes for the undoped and
hole-doped NiO$_2$ planes, respectively. The Ni$_5$O$_{16}$ cluster is embedded in an array of Ni 3$d^{9}$ sites which shift the effective on-site energy of the outer O orbitals due to the inter-site Coulomb energy~\cite{Hybertsen1990}.

Figure~\ref{fig:Heisenberg} is the low-energy spectrum mapping for the Ni5O16 cluster with five
holes in the insulating ground state of the undoped NiO$_2$ planes.  The spin-1/2
Heisenberg model
\begin{eqnarray}
  \label{eq:Heisenberg}
  H_J=\sum_{ij}J_{ij}\mathbf{S}_i\cdot\mathbf{S}_j,
\end{eqnarray}
with the nearest neighbor and next nearest neighbor exchange interactions
$J=28.6$~meV and $J'=0.4$~meV well reproduces the low-energy spectrum for the
three-band Hubbard model as shown in Fig.~\ref{fig:Heisenberg}. Therefore, we
obtain the effective Heisenberg model for the undoped NiO$_2$ planes.

\begin{figure}[b]
  \centering    
  \includegraphics[width=\columnwidth]{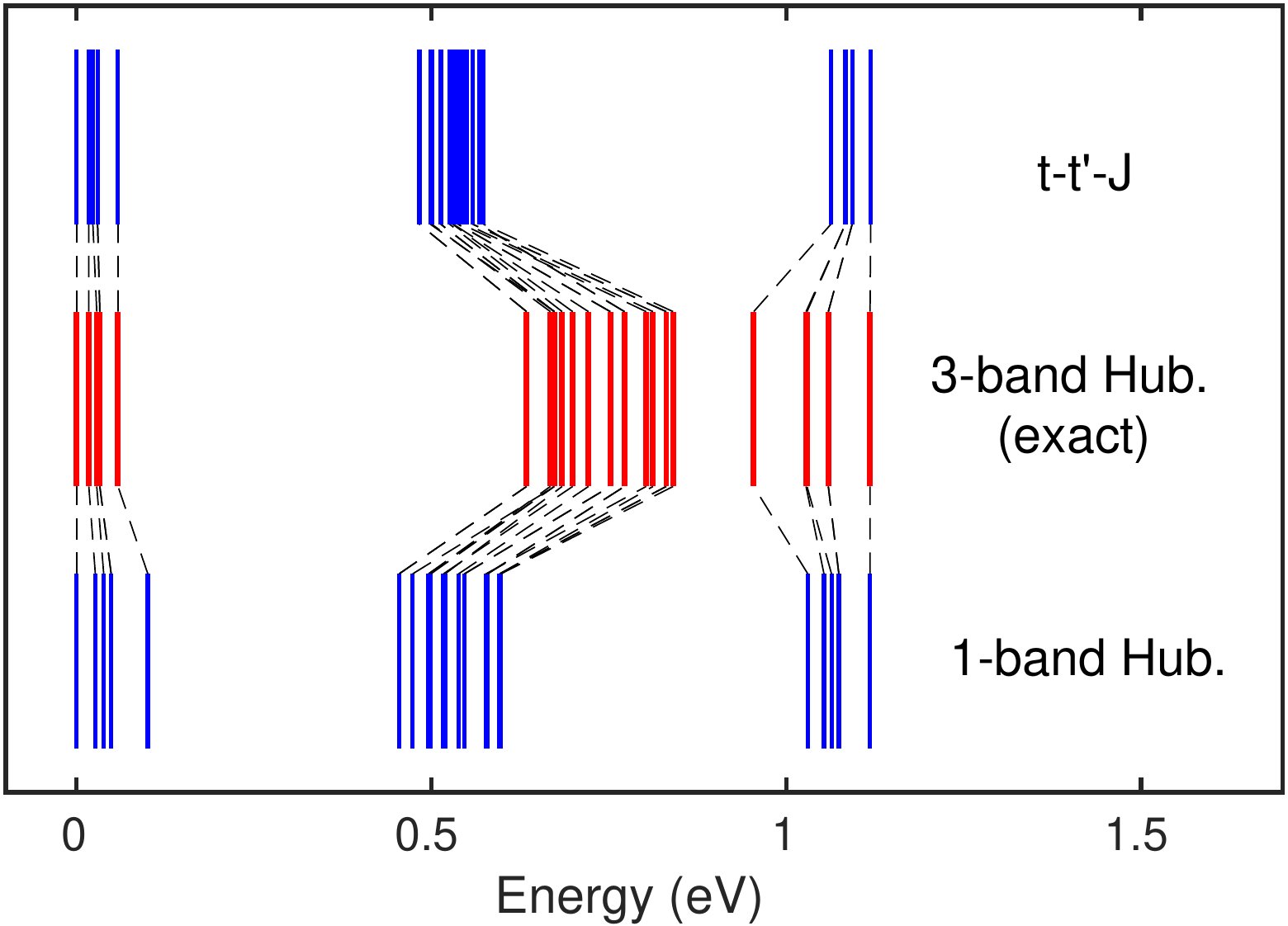}  
  \caption{The low-energy spectrum for Ni$_5$O$_{16}$ cluster mapping onto effective
    one-band $t$-$t'$-$J$ and Hubbard models.}
  \label{fig:tJ}
\end{figure}
For the hole-doped phase, the calculated low-energy spectra for six holes in the
Ni5O16 cluster is shown in Fig~\ref{fig:tJ}. The chemical character of the hole
in the ground state of the doped NiO2 planes is  Ni (76\%) and O (24\%)
character, indicating that NdNiO2 is in the regime of a Hubbard-Mott insulator,
different from the cuprate where the doped holes primarily locate on
O~\cite{Zhang1988, Hybertsen1990}. The significant difference comes from a
smaller $U_d$, but a larger $\epsilon=\epsilon_p-\epsilon_d$ in NdNiO$_2$ than
those in La$_2$CuO$_4$.  

We map the three-band Hubbard model in Eq.~(\ref{eq:3band}) onto the single-band $t$-$t'$-$J$ model
\begin{eqnarray}
  \label{eq:HtJ}
  H_{tJ}=\sum_{ij\sigma}t_{ij}c_{i\sigma}^\dag c_{j\sigma}+\sum_{ ij }J_{ij}\mathbf{S}_i\cdot\mathbf{S}_j,
\end{eqnarray}
where $c_{i\sigma}$ is the electron operator on the Ni sites connected by the
solid ($t$ and $J$) and dashed ($t'$ and $J'$ ) lines in Fig.~\ref{fig:Ni5O16}.
We first fix $J=28.6$~meV and $J'=0.4$~meV as in the undoped case and then tune
the hopping parameters $t$ and $t'$. The suitable values for $t=265$~meV,
$t'=-21$ meV with the mapping spectra shown in Fig.~\ref{fig:tJ}. The signs for
$t,t'$ refer to hole notation. We notice that the low-energy spectra has the
spectrum width around 58~meV, approximately twice of the exchange strength
$J=28.6$~meV, in a good agreement with $t$-$J$ model calculations for the
cuprate, which show that, independent of the value of $t$, the dressing of the
hole moving in the antiferromagnetic back-ground reduces the quasi-particle
bandwidth to the twice of $J$~\cite{Dagotto1994}.

We also map the low-energy spectra onto one-band Hubbard model
\begin{eqnarray}
  \label{eq:hubbard}
  H_{1b}=\sum_{ ij\sigma }t_{ij}c_{i\sigma}^\dag c_{j\sigma}+U\sum_in_{i\uparrow}n_{i\downarrow},
\end{eqnarray}
with $t=254$~meV, $t'=-30$~meV and $U=6$~eV in Fig.~\ref{fig:tJ}. The one-band Hubbard model gives the similar charge-transfer gap as the three-band Hubbard model in the Ni$_5$O$_{16}$ cluster.

\section{Discussions and Conclusions}\label{sec:discussion}
From the above derivation of the effective one-band $t-t'-J$ model Hamiltonian,
we find that the physics of NiO$_2$ planes in Nd$_{1-x}$Sr$_x$NiO$_2$ is a doped Mott
insulator, even further away from a conventional Fermi liquid than the
superconducting cuprates. The undoped CuO$_2$ plane in the cuprate is a
charge-transfer type Mott insulator, but the undoped NiO$_2$ plane in the
nickelate is a Hubbard Mott insulator. Both the NiO$_2$ and CuO$_2$ planes have
the same effective one-band $t$-$t'$-$J$ model Hamiltonians, and they have the similar
physics of the superconductivity.

The discovering of the superconductivity in the nickelate oxide~\cite{Li2019}
certainly motivates the studies of the effective Hamiltonian in this paper.
However, two recent experimental papers reported non-superconductivity results
in Nd$_{1-x}$Sr$_x$NiO$_2$~\cite{Li2019a,Zhou2019}, getting the presence of
superconductivity into controversy. The existence of the superconductivity in
the doped nickelate oxides is a crucial issue and needs further exploring in the
experiments. In this paper, our study is the consequence of a quantum-chemical
description of Nd$_{1-x}$Sr$_x$NiO$_2$; however, we cannot provide direct
theoretical implication of the existence of the superconductivity. The effective
Hamiltonian in this work provides essential support for, and constraint on, 
models to describe the low-energy physics of the nickelate oxides, regardless of
the presence of the superconductivity. In our recent experimental
work~\cite{Fu2019}, we perform Raman scattering on  NdNiO$_2$ single crystal and
measure the Heisenberg superexchange strength $J=25$~meV from the two-magnon peak,
in a good agreement with our present work. Although the current situation is not
clear, we hope that our work will help us understand the electronic structure of
nickelate oxides.


In conclusion, we have explicitly derived a single-band effective $t$-$t'$-$J$
model Hamiltonian for Ni-O based compounds starting from a three-band model
based on the density functional theory.

\emph{Note added --} At the stage of finishing the manuscript, we noticed
related theoretical works for nickelates~\cite{Botana2019, Sakakibara2019,
  Hirsch2019, Jiang2019, Wu2019, Nomura2019, Gao2019, Ryee2019}. The paring of
hole carriers in the oxygen $p\pi$ orbitals is discussed in
Ref.~\cite{Hirsch2019}.  First-principles simulations within GGA have been worked
out in
Refs.~\cite{Botana2019,Ryee2019,Gao2019,Nomura2019,Wu2019,Sakakibara2019}. DMFT
is carried out in Ref.~\cite{Ryee2019}, and RPA analysis of the pairing symmetry
is done in Ref.~\cite{Wu2019, Sakakibara2019}. The hoping parameters for the
three-band Hubbard model are (implicitly or explicitly) given from the Wannier
fitting in
Refs.\cite{Botana2019,Sakakibara2019,Wu2019,Nomura2019,Gao2019,Ryee2019}, with
similar  values to our work. The hoping parameters ($t$ and $t'$) of the effective Ni
3$d_{x^2-y^2}$ band are also given in
Refs.~\cite{Botana2019,Wu2019,Nomura2019,Gao2019} with $t\sim 370$~meV and
$t'\sim -100$~meV, larger than the values $t=265$~meV and $t'=-21$~meV
renormalized by correlations in our
work. The exchange interaction  is estimated as $J=100$~meV in
Ref.~\cite{Wu2019}, larger than our values $J=28.6$ meV. The Hubbard-Mott scenario is also proposed in
Ref.~\cite{Jiang2019} and the charge-transfer gap is estimated $\epsilon=7\sim
9$~eV, larger than our value $\epsilon=4.2$~eV. In the Ref.~\cite{Jiang2019},
the $S=1$ Ni$^{2+}$ state is proposed when the hope is doped into the NiO$_2$
planes, different from the $S=0$ Ni$^{2+}$ state in our work. We notice that if
the charge-transfer gap is taken as the value we use in the work,
$\epsilon=4.2$~eV, the $S=0$ Ni$^{2+}$ state is favored according to the
calculation in Ref.~\cite{Jiang2019}.

\section{acknowledgements}
 J.W.M thanks J.S. Chen
 for the help of drawing Fig.~1. J.W.M was partially supported by the program for Guangdong Introducing
 Innovative and Entrepreneurial Teams (No.~2017ZT07C062). F.Y was supported by
 National Science Foundation of China (No.~11774143). The exact diagonalization
 is performed by $\mathcal{H}\Phi$ package\cite{Kawamura2017}.

 \appendix

 \section{Supplementary DFT results for
   Nd$_{1-x}$Sr$_x$NiO$_2$}\label{sec:NdNiO}
 In this part, we provide the supplementary DFT results for
 Nd$_{1-x}$Sr$_x$NiO$_2$. The main purpose of the supplementary results is
 two-folded: (a) check the validity of the Nd pseudopotential with the
 core-level $4f$ electrons; (b) compare the GGA and SCAN band structures to the
 HSE06 results in the main text. The spin-orbital coupling is also checked in
 this section.

 \begin{figure}[b]
   \centering
   \includegraphics[width=\columnwidth]{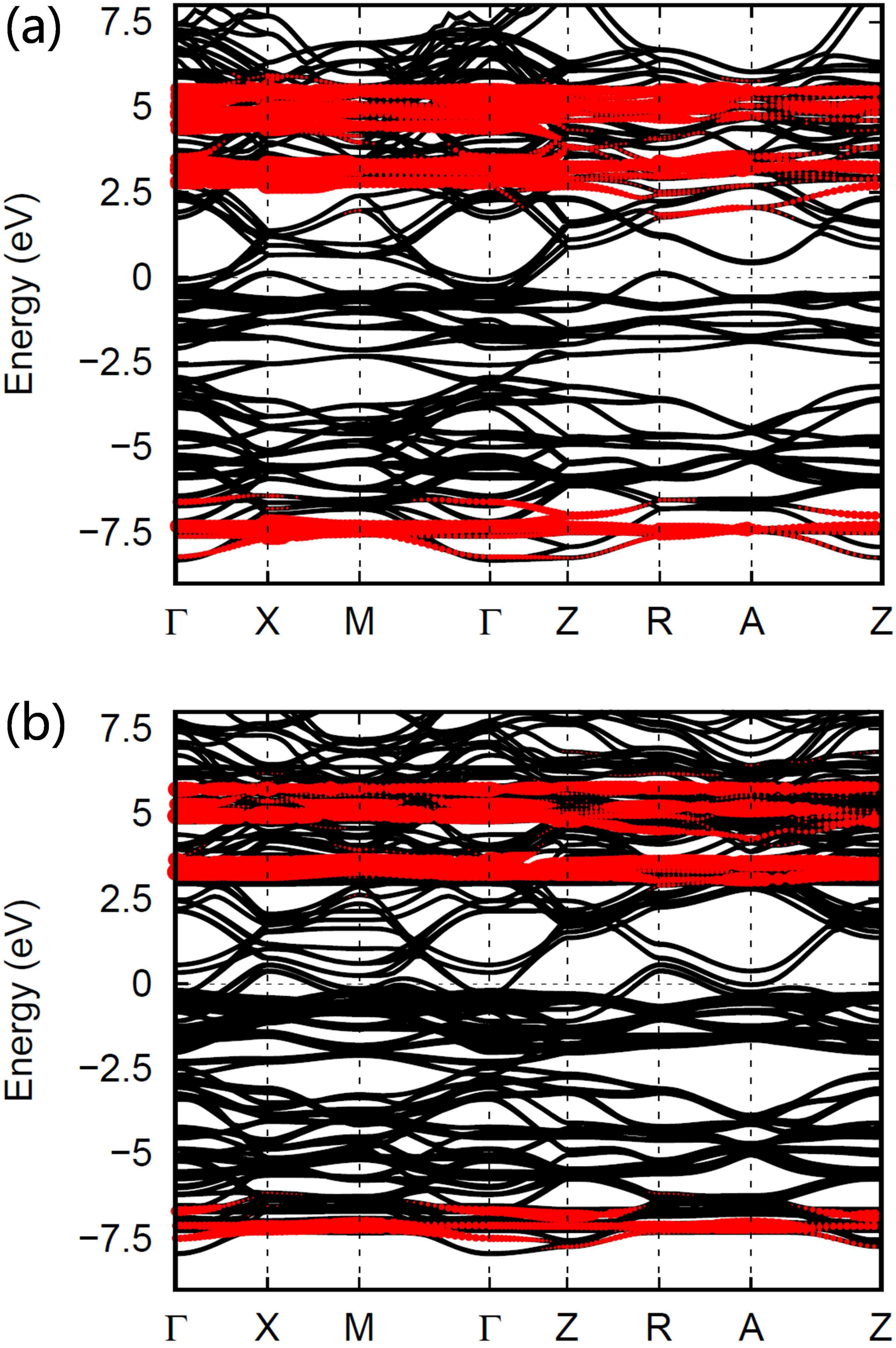}
   \caption{GGA band structures of G-type AFM states for a  supercell in (a) NdNiO2 and (b) Nd0.75Sr0.25NiO2 calculated with 4f valence electrons pseudopotential within GGA+U (U4f = 10 eV) scheme with SOC. The notations of $\Gamma$, X, M, Z, R and A are in the folded magnetic Brillouin zone. The projected band structures of f orbitals (red) in Nd are also shown. The Fermi level is set at 0 eV.}
   \label{fig:figureS3}
 \end{figure}
\subsection{GGA+U calculations for the Nd $4f$ valence electron pseudopotential}\label{sec:Ndf}
As Nd$^{3+}$ in Nd$_2$CuO$_4$~\cite{Casalta1998} and Ho$^{3+}$ in
HoNiO$_3$~\cite{Fernandez-Diaz2001}, Nd$^{3+}$ has the local $4f$ moment far
below the Fermi energy level $E_F$. In the main text, we treat the $4f$
electrons in Nd$^{3+}$ as the core-level electrons in the Nd pseudopotential. In
this subsection, we verify the validity of this treatment. 

Figure~\ref{fig:figureS3} are the band structure of G-type AFM states for
NdNiO$_2$ and Nd$_{0.75}$Sr$_{0.25}$NiO$_2$ calculated with $4f$ valence
electron pseudopotential within GGA+U scheme. Without the U term, the $4f$
electrons form very localized bands near the Fermi level. In the G-type AFM
states, the local moments of $4f$ electrons are also AFM within the same Nd
plane and between Nd planes along the $c$ direction. We take the on-site
interaction for $4f$ electrons $U_{4f}=10$~eV in the GGA+U calculation which
splits the $4f$ bands above and below the Fermi energy level $E_F$. During the
calculations, we also include the spin-orbital couplings. 

According to Fig.~\ref{fig:figureS3}, we can see that the $4f$ electron states
couple to Nd 5$d$ bands, however, doesn't significantly change the band
structures near the Fermi level. Thus we can treat 4f orbitals as the
core-level electrons in the Nd pseudopotential.

\subsection{GGA and SCAN band structures}\label{sec:gga}
\begin{figure*}[t]
  \centering
  \includegraphics[width=2\columnwidth]{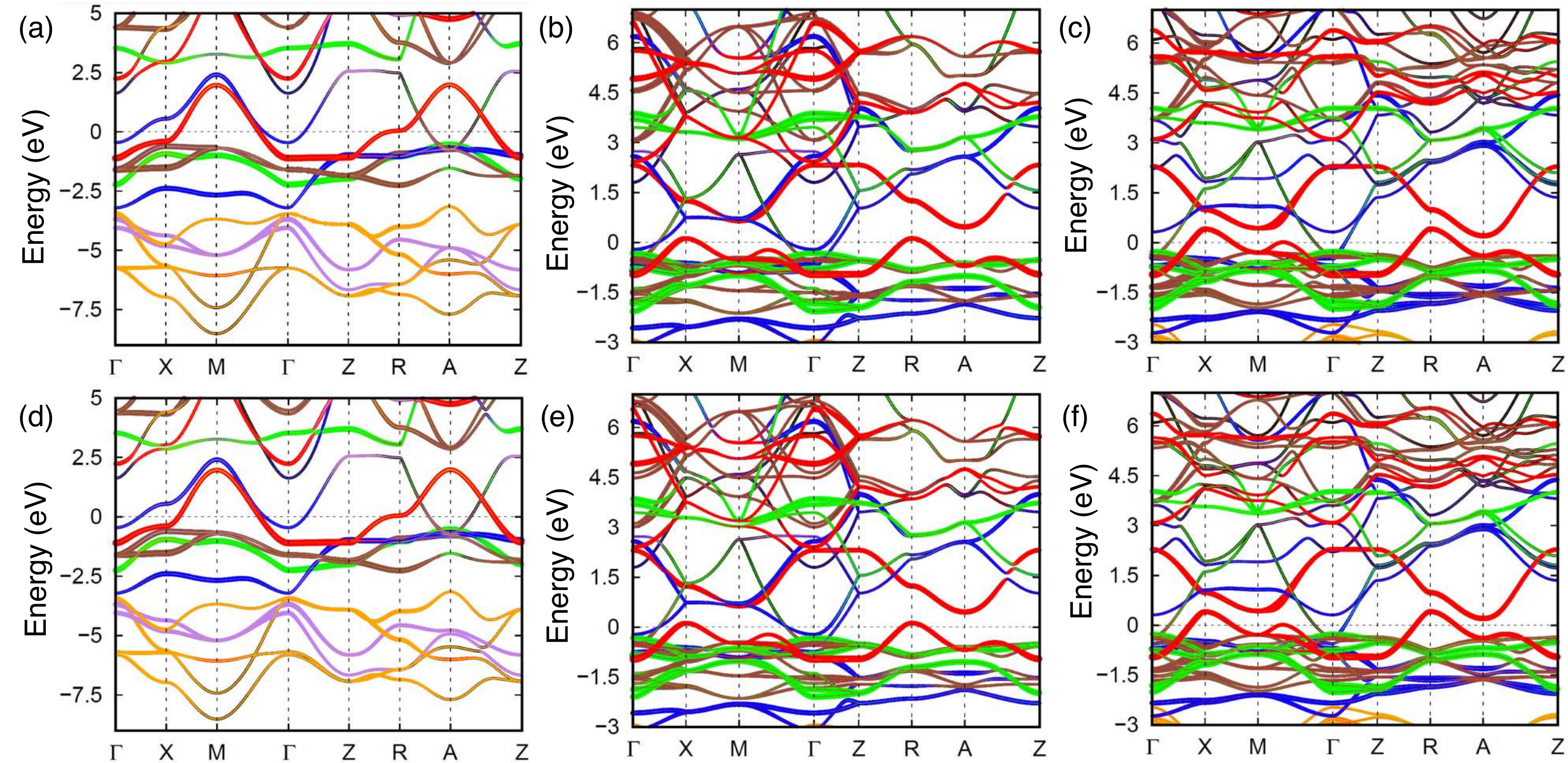}
  \caption{GGA band structures of Nd$_{1-x}$Sr$_{x}$NiO$_2$. (a),  (b), and (c)
    correspond to Fig.~\ref{fig:figure1}~(b) (the non-magnetic state for
    NdNiO$_2$), Fig.~\ref{fig:band_afm}~(a) (the G-type AFM state for NdNiO$_2$) and
    (b) (the G-type AFM state for Nd$_{0.75}$Sr$_{0.25}$NiO$_2$), respectively,
    without SOC.  (d), (e), and (f) are the results with turning on SOC. }
  \label{fig:figureS4}
\end{figure*}

\begin{figure}[t]
  \centering
  \includegraphics[width=\columnwidth]{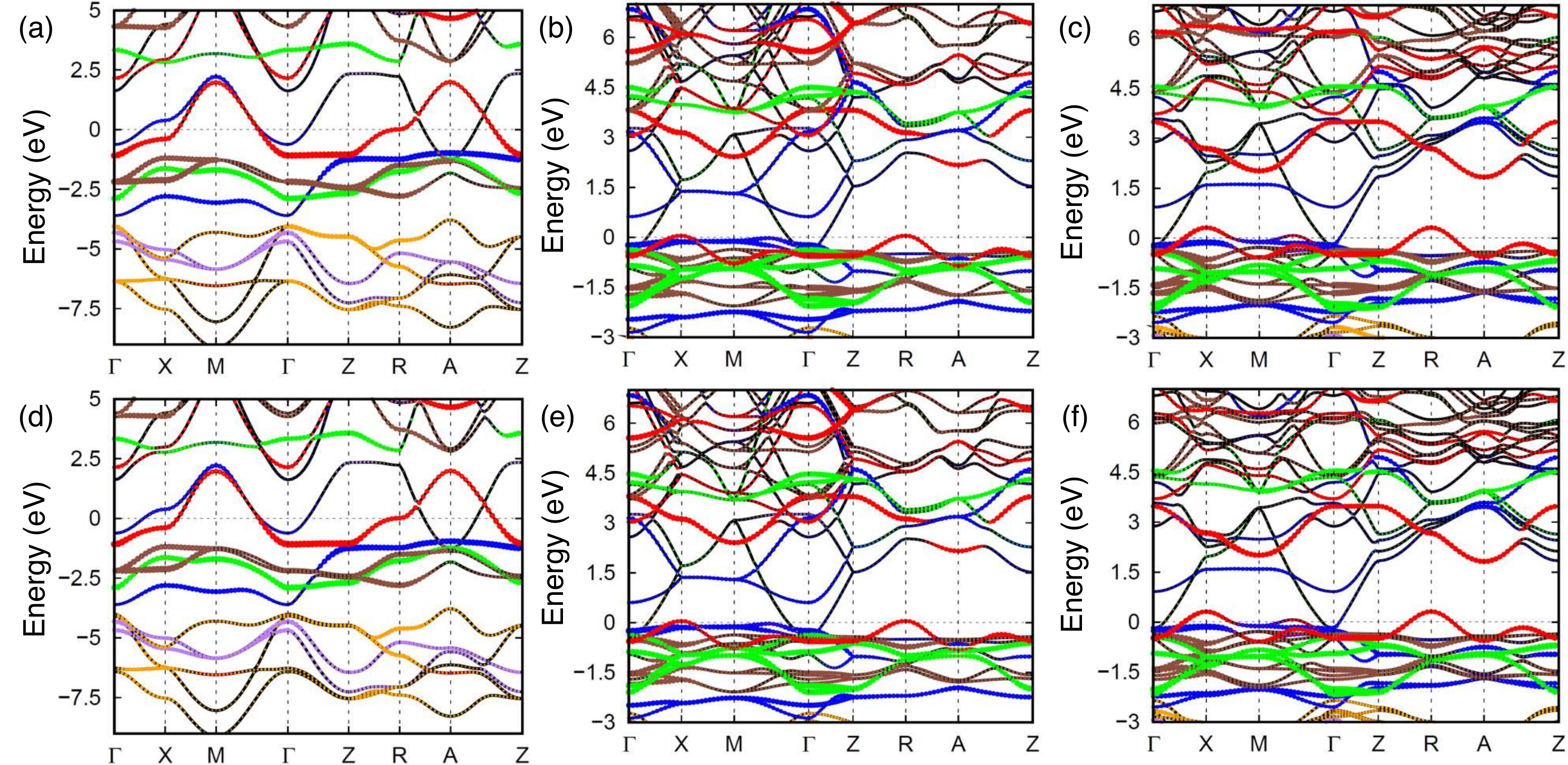}
  \caption{SCAN band structures of Nd$_{1-x}$Sr$_x$NiO$_2$, similar to Fig.~\ref{fig:figureS4}.}
  \label{fig:figureS5}
\end{figure}

In this subsection, we present the band structures for Nd$_{1-x}$Sr$_{x}$NiO$_{2}$
within GGA and SCAN functionals without/with spin-orbital couplings in
Fig.~\ref{fig:figureS4} and Fig.~\ref{fig:figureS5}, respectively. Again, once
soc is taken into account, there is no significant change in the band
structures. 
The main features of SCAN band structures are very close to those in HSE06 band
structures.

\section{HSE06 results of LaNiO$_2$}
\begin{figure}[t]
  \centering    
  \includegraphics[width=\columnwidth]{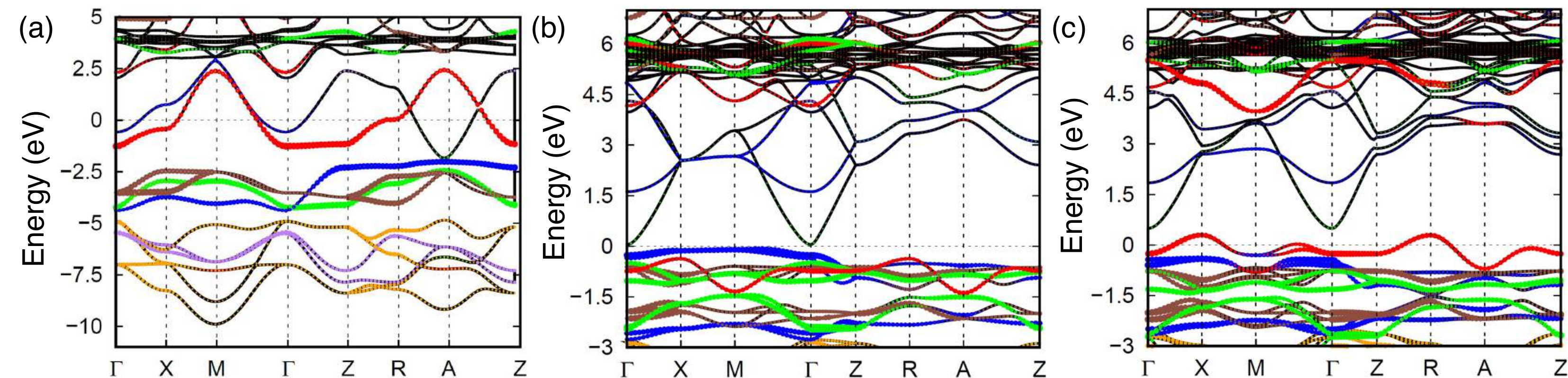}  
  \caption{HSE06 band structures of La$_{1-x}$Sr$_x$NiO$_2$ without SOC. (a) non-magnetic state for
    LaNiO$_2$,  (b) G-type AFM state for LaNiO$_2$, and (c) G-type AFM state for
    La$_{0.75}$Sr$_{0.25}$NiO$_2$, 
    corresponding to Fig.~\ref{fig:figure1}~(b) , Fig.~\ref{fig:band_afm}~(a)  and
    (b) , respectively.}
  \label{fig:S1}
\end{figure}

The LDA result for LaNiO$_2$ has been studied previously~\cite{Lee2004}. In this
subsection, we implement the HSE06 hybrid functional to calculated band structures of the non-magnetic state
in LaNiO$_2$ and G-type AFM states for a  supercell in LaNiO$_2$ and
La$_{0.75}$Sr$_{0.25}$NiO$_2$ in Fig.~\ref{fig:S1}. The band structures are very
similar to the results of NdNiO$_2$, implying the generic electronic structures in the
RNiO$_2$ family.

\clearpage
\onecolumngrid

\twocolumngrid
\FloatBarrier
\bibliography{../HTS}
\end{document}